\documentclass[fleqn,10pt]{wlscirep}
\usepackage{amsfonts,amsmath,bm,multirow}

\usepackage{graphicx}
\usepackage{xcolor}
\usepackage{epstopdf}
\usepackage{dsfont}
\usepackage{tabularx}
\usepackage{hf-tikz}
\usepackage{physics}
\usepackage{hyperref}
\usepackage{placeins}

\usepackage[utf8]{inputenc}
\usepackage[T1]{fontenc}
\usepackage{siunitx} 
\usepackage{textcomp}

\newcommand{\kp}{\bm{k}\!\vdot\!\bm{p}}
\newcommand{\rr}{\bm{r}}
\newcommand{\kk}{\bm{k}}

\newcommand{\mB}{\mu_{\mathrm{B}}}

\definecolor{bred}{HTML}{e31a1c}
\definecolor{bgreen}{HTML}{33a02c}
\definecolor{bblue}{HTML}{1f78b4}

\definecolor{armygreen}{rgb}{0.29, 0.33, 0.13}

\renewcommand{\cp}{{\mathrm{c.p.}}}

\begin{document}
	\title{Electron g-factor in nanostructures: continuum media and atomistic approach}
	\author[1,*]{Krzysztof Gawarecki}
	\affil[1]{Department of Theoretical Physics, Faculty of Fundamental Problems of Technology, Wroc\l aw University of Science and Technology, Wybrze\.ze Wyspia\'nskiego 27, 50-370 Wroc\l aw, Poland}
	\author[2]{Micha{\l} Zieli{\'n}ski}
	\affil[2]{
	Institute of Physics, Faculty of Physics, Astronomy and Informatics, Nicolaus Copernicus University, ul. Grudziadzka 5, 87‑100 Toru\'n, Poland}
	\affil[*]{Krzysztof.Gawarecki@pwr.edu.pl}
	
	\begin{abstract}
    We report studies of $\kk$-dependent Land\'e $g$-factor, performed by both continuous media approximation $\kp$ method, and atomistic tight-binding sp$^3$d$^5$s$^*$ approach.
    We propose an effective, mesoscopic model for InAs that we are able to successfully compare with atomistic calculations, for both very small and very large nanostructures, with a number of atoms reaching over 60 million.
    Finally, for nanostructure dimensions corresponding to near-zero $g$-factor we report electron spin states anti-crossing as a function of system size, despite no shape-anisotropy nor strain effects included, and merely due to breaking of atomistic symmetry of cation/anion planes constituting the system.
	\end{abstract}
	
	\maketitle
	
	\section{Introduction}
	\label{sec:intr}
	The measurement and control of a confined electron spin is an important branch of studies on semiconductor nanostructures. The linear response of an electron to magnetic field
	can be conveniently described with a single parameter -- Lande's $g$-factor~\cite{Hentschel2009}. Accurate modelling of $g$-factors in nanostructures must 
	be able to reproduce two apparent extremes: free-electron Dirac equation value of $g \approx 2$ for small nanostructures with a strong, confining potential, and
	reproduce bulk $g$-factor significantly re-normalized in solids by the spin-orbit interaction~\cite{Roth1959,Kiselev1998}.
	In case of indium arsenide (InAs), which is fundamentally important building component of self-assembled and nanowire quantum dots, the bulk $g$-factor reaches a substantial value of $-14.7$~\cite{Konopka1967}. A treatment aiming to model $g$-factor in nanostructures must therefore be able to address these two asymptotic cases, as well as work well in an intermediate regime of mesoscopic dimensions.
	Moreover, the real-space description of $g$-factor should have its counterpart in the $\kk$-space.
	
	Despite the value of band-edge electron $g$-factor in bulk semiconductors can be predicted by the well established Roth-Lax-Zwerdling formula~\cite{Roth1959}, recent years bring development to the theoretical understanding of systems with the reduced dimensionality. In fact, many theoretical works explores $g$-factors in nanostructures, including quantum wells~\cite{Ivchenko1997,Gradl2014,Gradl2018} and quantum dots of various types (e.g. spherical~\cite{Kiselev1998,Rodina2003,Tadjine2017}, nanowire-embedded\cite{Schroer2011} and self-assembled~\cite{Andlauer2009,Gawarecki2018a,Mielnik-Pyszczorski2018,Kahraman2020} systems). 
	The works utilize a variety of approaches, including multiband $\kp$ methods~\cite{Andlauer2009,Gawarecki2018a,Mielnik-Pyszczorski2018}, tight-binding models~\cite{Schrier2003,Tadjine2017}, and empirical pseudopotential framework~\cite{Kahraman2020}. The information about relations between the models, their assumptions and accuracy is useful from the theoretical point of view. 

	In this work, we utilize a general theory of the magnetic-field dependence of the Bloch states~\cite{Wozniak2020} applied in the frameworks of the eight-band $\kp$ and the sp$^3$d$^5$s$^*$ tight-binding method. We calculate the $\kk$-dependent effective $g$-factors ($g(\kk)$) for conduction band (CB) states in bulk InAs and demonstrate a relatively good agreement between the methods. 
	Based on the bulk $\kk$-dependence we introduce a simplified, mesoscopic model. We compare this model with $\kp$ and atomistic tight-binding results, by calculating a size-dependence of $g$-factor for a cubic InAs nanostructure with a size varying from a single lattice constant up to over 120~nm leading to challenging multi-million atom simulations. Importantly, computationally cost-effective model based on the $g(\kk)$, gives size-dependent $g$-factor values inter-mediating between results of atomistic tight-binding and $\kp$ method based on the continuous media approximation.
	
	Finally, we focus on near-zero region of $g$-factor values, i.e. nanostructure spatial dimension where $g$-factor changes sign. Systems with a vanishing $g$-factor are particularly important in the context of inducing strong coupling between electron and nuclear spin baths, 
	studies of spin textures, and possible further application in quantum information and communications~\cite{Kahraman2020,Kosaka2001}.
	We show however there mere difference in atomic scale arrangement of ionic layers of a nanostructure, i.e. low atomistic symmetry, leads to an apparent anti-crossing of electron spin states, and prevents $g$-factor from vanishing. This is also interesting since we do not consider hole $g$-factors strongly affected by band mixing effects, and expected to reveal strong anisotropy, but electron states dominated by contribution for isotropic $s$ atomic orbitals.

	\section{Bulk material}
	\label{sec:bulk}
	The first part of this Section contains a general introduction devoted to the linear response theory describing the Bloch state g-factors. Then, we use this approach to calculate $g(\kk)$ in terms of the eight-band $\kp$ and  sp$^3$d$^5$s$^*$ tight-binding models. In the last part, we present numerical results for $g(\kk)$ in the InAs bulk crystal, and introduce a simple mesoscopic model
	$g_{\mathrm{eff}}(\kk)$ for InAs.
	
    \subsection{Introduction: magnetic-field dependence of the Bloch states}	
    The electron Land\'e g-factor in a bulk can be calculated within the linear response theory~\cite{Kiselev1998,eissfeller12} or using the Berry phase formula~\cite{Thonhauser2005,Ceresoli2010}. In the first approach, magnetic field enters as a perturbation (in the first order) to the Hamiltonian at $\bm{B}=0$. For the magnetic field oriented along the $z$ direction, this reads~\cite{Kiselev1998,eissfeller12}
	$$
		g_\varphi = g_0 + \frac{1}{\hbar} \qty( \mel{\varphi \uparrow}{\hat{L}_z}{ \varphi \uparrow} - \mel{\varphi \downarrow}{\hat{L}_z}{ \varphi \downarrow} )
	$$
	where $g_0 \approx 2$, the states $\ket{\varphi \uparrow / \downarrow} = \ket{\varphi} \otimes \ket{ \uparrow / \downarrow} $ (the orbital part is represented by $\ket{\varphi}$ and  $\ket{ \uparrow / \downarrow}$ is a spinor), and $\hat{L}_z$ is the axial component of the angular momentum operator.
    
	According to the Bloch's theorem, the states in a bulk crystal can be written as
	$$
	\ket{\Psi_{\alpha\kk}} = e^{i \kk \rr} \ket{\alpha \kk},
	$$
	where $\alpha$ is the band index (with the corresponding energy $E_{\alpha}(\kk)$). The $\hat{L}_z$ matrix elements can be calculated by~\cite{Wozniak2020}
	\begin{align}
	\label{eq:Lz}
	&\mel{\Psi_{\alpha\kk}}{\hat{L}_z}{\Psi_{\alpha\kk}} \equiv L_{z,\alpha}(\kk) =  \bra{\Psi_{\alpha\kk}} \hat{x} \hat{p}_y - \hat{y} \hat{p}_x \ket{\Psi_{\alpha\kk}} \nonumber \\
	&=   \sum_{\beta} \int \dd{\kk'} \Big \{ \bra{\Psi_{\alpha \kk}} \hat{x} \ketbra{\Psi_{\beta \kk'}} \hat{p}_y \ket{\Psi_{\alpha \kk}}  - \bra{\Psi_{\alpha \kk}} \hat{y} \ketbra{\Psi_{\beta \kk'}} \hat{p}_x \ket{\Psi_{\alpha \kk}} \Big \} \nonumber \\
	&=   \sum_{\beta \neq \alpha}  \Big \{ R_{x,\alpha\beta}(\kk)  P_{y,\beta\alpha}(\kk) - R_{y,\alpha\beta}(\kk)  P_{x,\beta\alpha}(\kk) \Big \},
	\end{align}
	where $P_{i,\alpha\beta}(\kk) \equiv \mel{\Psi_{\alpha \kk}}{ \hat{p}_i }{\Psi_{\beta \kk}} $ and the position operator matrix elements are calculated from~\cite{LewYanVoon1993}
	\begin{equation}
	\label{eq:pos}
	R_{i,\alpha\beta}(\kk) =  \frac{-i \hbar}{m_0} \frac{1}{E_\alpha(\kk)-E_\beta(\kk)} P_{i,\alpha\beta}(\kk).
	\end{equation}

	\subsection{Conduction band g-factors}
	In presence of the spin-orbit interaction, the spin operator $\hat{S}_z$ does not commute with the Hamiltonian and states described by some nominal spins, contains admixtures of the opposite orientation\cite{Bir1974}. Hence, a state can not be (generally) expressed as a single product of the orbital and spin part. 
	However, for the (mainly $s$-type) conduction band in InAs, the effect of the SO coupling is relatively weak in the vicinity of $\kk=0$.
	One can build superpositions for which $\expval{S_z} \approx \pm {\hbar}/{2}$. To this end, the $\hat{S}_z$ operator is diagonalized~\cite{eissfeller12}. In the basis of the CB Bloch states this gives
	$$
	\ket{ \widetilde{\Psi}_{\mathrm{cb};\kk}, \pm \frac{1}{2}} \approx a_{\pm} \ket{\Psi_{\alpha\kk}} + b_{\pm} \ket{\Psi_{\alpha'\kk}},
	$$
	where $a_{\pm}, b_{\pm}$ are coefficients found from the diagonalization and $\ket{\Psi_{\alpha\kk}}, \ket{\Psi_{\alpha'\kk}}$ is a pair of CB states which are slightly splitted ($E_{\alpha}(\kk) \approx E_{\alpha'}(\kk)$) due to the Dresselhaus spin-orbit coupling.
	 The effective bulk CB g-factor calculated between such configurations is given by 
	\begin{align*}
	\label{eq:gf}
	\begin{split}
	g_{\mathrm{cb}}(\kk) = ~& g_0 + \frac{1}{\hbar} \Bigg ( \mel{  \widetilde{\Psi}_{\mathrm{cb};\kk}, \frac{1}{2}}{\hat{L}_z}{  \widetilde{\Psi}_{\mathrm{cb};\kk}, \frac{1}{2}} - \mel{  \widetilde{\Psi}_{\mathrm{cb};\kk}, -\frac{1}{2} }{\hat{L}_z}{  \widetilde{\Psi}_{\mathrm{cb};\kk}, - \frac{1}{2}} \Bigg ).
	\end{split}
	\end{align*}	
	
	The Bloch states can be expressed by
	$$
	\ket{\Psi_{\alpha\kk}} = e^{i \kk \rr} \sum_m c^{(\alpha)}_m(\kk) \ket{m},
	$$
	where $c^{(\alpha)}_m(\kk)$ are coefficients, $\ket{m}$ are states at a chosen $\kk_0$ point (in the $\kp$ approach) or atomic orbitals (in TB models). The index $m$ carries information about the orbital and spin part \mbox{$\{ \varphi, \uparrow / \downarrow \}$}.
	Then the momentum matrix elements can be written as
	\begin{equation}
	\label{eq:pk}
	P_{i,\alpha\beta}(\kk) = \sum_{n,m} c^{(\alpha)*}_n(\kk) \, c^{(\beta)}_m(\kk) \, \widetilde{P}_{i,nm}(\kk),
	\end{equation}
	where $\widetilde{P}_{i,nm}(\kk)$ can be calculated using the Hellmann-Feynman theorem\cite{Feynman1939, LewYanVoon1993, eissfeller12}
	\begin{equation}
	\label{eq:hf}
	\widetilde{P}_{i,nm}(\kk) \approx \frac{m_0}{\hbar} \pdv{H_{nm}(\kk)}{k_i},
	\end{equation}
	where $H_{nm}(\kk) \equiv \mel{n}{H(\kk)}{m}$ are the bulk Hamiltonian (at $\bm{B} = 0$) matrix elements.	
	\subsection{Eight band k.p model}
	\label{sec:bulk_kp}
	In the eight-band $\kp$ model, the invariant expansion of the bulk Hamiltonian (at $\bm{B} = 0$) is given by\cite{Luttinger1955,Trebin1979,Winkler2003}
	\begin{subequations}
		\begin{align*}
		H_{\mathrm{6c6c}} &= E_{g} + \frac{\hbar^{2}}{2m_0} (k_x^{2} + \cp ),\\
		H_{\mathrm{8v8v}} &= -\frac{\hbar^{2}}{2m_{0}} \left \{ \gamma'_{1}  k_x^{2} - 2 \gamma'_{2}  \left (  J^{2}_{x}  - \frac{1}{3} J^{2}  \right ) k^{2}_{x}  \right . \nonumber \\
		&\phantom{=} - 4 \gamma'_{3} \{J_{x},J_{y}\}  \{k_{x},k_{y}\} + \cp \bigg \}, \nonumber\\
		H_{\mathrm{7v7v}} &= -\Delta_{0} - \frac{\hbar^{2} }{2m_{0}} \gamma'_{1} (k_x^{2} + \cp ), \nonumber \\
		H_{\mathrm{8v7v}} &= \frac{3 \hbar^{2}}{m_{0}} \left[ \gamma'_{2} T^\dagger_{xx} k^{2}_{x} + 2 \gamma'_{3} T^\dagger_{xy} \{ k_{x} , k_{y} \} + \cp \right ], \nonumber \\
		H_{\mathrm{6c8v}} &= \sqrt{3} P_0 \bm{T} \cdot \kk,\\ 
		H_{\mathrm{6c7v}} &= - \frac{1}{\sqrt{3}}  P_0 \bm{\sigma} \cdot \kk,
		\end{align*}
	\end{subequations}	
	where $6\mathrm{c}$, $8\mathrm{v}$, and $7\mathrm{v}$ are related to the conduction- and valence band blocks (this notation corresponds to the irreducible representations of the $T_\mathrm{d}$ point group), $E_\mathrm{g}$ is the energy gap, $\Delta_0$ is a parameter accounting for the spin-orbit interaction in the valence band, $\gamma'_{1-3}$ are the reduced (by subtracting the contributions coming from the $6\mathrm{c}$ band block) Luttinger parameters, $P_0$ is a parameter proportional to the interband momentum matrix element; $\sigma_i$ are the Pauli matrices, while the explicit definitions of matrices $J_i$, $T_i$, and $T_{ij}$ are provided in Refs.\cite{Winkler2003,Trebin1979}. We use the values of material parameters  given in the Appendix of Ref.~\cite{Gawarecki2018a}, except for the $P_0 = \sqrt{ E_\mathrm{p}  \hbar^{2} / (2 m_{0} )}$ which is taken with $E_\mathrm{p} = 21.2$. The remote band contributions to the electron effective mass are neglected. We also neglect the Dresselhaus SO terms and small $k$-linear terms related to the inversion asymmetry (the $C_k$ parameter).
	Finally, the Hamiltonian is diagonalized (at each considered $\kk$, separately), which gives the $c^{(\alpha)}_m(\kk)$ coefficients.
	
	To calculate the $g_{\mathrm{cb}}(\kk)$, one needs to find the momentum and position matrix elements constituting the $L_{z,cb}(\kk)$ (see Eq.~\ref{eq:Lz}).
	The momentum block matrices are calculated using Eq.~\ref{eq:hf}
	\begingroup
	\allowdisplaybreaks
	\begin{subequations}
	\begin{align*}
	\widetilde{P}_{x,\mathrm{6c6c}}(\kk) &= \frac{m_0}{m'_{\mathrm{e}}} k_x,\\
	\widetilde{P}_{x,\mathrm{8v8v}}(\kk) &= - \hbar \bigg \{  \gamma'_{1}  k_x - 2 \gamma'_{2}  \left (  J^{2}_{x}  - \frac{1}{3} J^{2}  \right ) k_{x}   - 2 \gamma'_{3} \big ( \{J_{x},J_{y}\} k_{y} + \{J_{z},J_{x}\} k_{z} \big ) \bigg \}, \nonumber\\
	\widetilde{P}_{x,\mathrm{7v7v}}(\kk) &= - \hbar  \gamma'_{1} k_x, \nonumber \\
	\widetilde{P}_{x,\mathrm{8v7v}}(\kk) &= \frac{m_0}{\hbar} \frac{3 \hbar^{2}}{m_{0}} \left[ 2 \gamma'_{2} T^\dagger_{xx} k_{x} + 2 \gamma'_{3} (T^\dagger_{xy} k_{y} + T^\dagger_{zx} k_{z}) \right ], \nonumber \\
	\widetilde{P}_{x,\mathrm{6c8v}}(\kk) &= \frac{m_0}{\hbar} \sqrt{3} P_0 T_x,\\ 
	\widetilde{P}_{x,\mathrm{6c7v}}(\kk) &= - \frac{m_0}{\hbar} \frac{1}{\sqrt{3}}  P_0 \sigma_x,
	\end{align*}
	\end{subequations}
	\endgroup
	where $\widetilde{P}_y(\kk)$ and $\widetilde{P}_z(\kk)$ matrices can be found from cyclic permutations. Then, the $P_{i,\alpha\beta}(\kk)$ are calculated from Eq.~\ref{eq:pk}, while the $R_{i,\alpha \beta}(\kk)$ result from Eq.~\ref{eq:pos}. 
	
	For $\kk=0$ the problem is reduced to the well known case of the band-edge g-factor~\cite{Roth1959,Kiselev1998}, where the states in electron Zeeman doublet are purely $S$-type and (at $\bm{B}=0$) belong to the two-dimensional $\Gamma_{6{\mathrm{c}}}$ representation, giving
		\begin{align*}
	\begin{split}
	L_{z,\Gamma_{6\mathrm{c}} \Gamma_{6\mathrm{c}}}(0) =&  \frac{-i \hbar}{m_0} \sum_{\beta \neq  6\mathrm{c}} \Big \{ \frac{P_{x,\Gamma_{6\mathrm{c}} \Gamma_\beta}(0) P_{y,\Gamma_\beta \Gamma_{6\mathrm{c}}}(0)}{E_{\mathrm{g}}-E_{\Gamma_\beta}(0)}  -  \frac{P_{y,\Gamma_{6\mathrm{c}} \Gamma_\beta}(0) P_{x,\Gamma_\beta \Gamma_{6\mathrm{c}}}(0)}{E_{\mathrm{g}}-E_{\Gamma_\beta}(0)} \Big \} \\
	=& \frac{1}{3} \hbar E_\mathrm{p} \left [ \frac{1}{E_\mathrm{g}+\Delta_0} - \frac{1}{E_\mathrm{g}} \right ] \sigma_z.
	\end{split}
	\end{align*}
	%
	The electron g-factor is then given by
	\begin{align*}
	\begin{split}
	{g}(0) = ~& g_0 + \frac{1}{\hbar} \mel{ \Gamma_{6\mathrm{c}}, \frac{1}{2}}{\hat{L}_z}{ \Gamma_{6\mathrm{c}},  \frac{1}{2}} - \frac{1}{\hbar} \mel{ \Gamma_{6\mathrm{c}}, - \frac{1}{2}}{\hat{L}_z}{  \Gamma_{6\mathrm{c}}, - \frac{1}{2}} \\ = ~& g_0 + \frac{2}{3} E_\mathrm{p} \left [ \frac{1}{E_\mathrm{g}+\Delta_0} - \frac{1}{E_\mathrm{g}} \right ], 
	\end{split}
	\end{align*}	
	which reproduces the well-known Roth-Lax-Zwerdling formula\cite{Roth1959}. 

	\subsection{Tight-binding model}	
	The tight-binding model~\cite{Slater1954,Chadi1977,Korkusinski2008,Yu2010} Hamiltonian can be written in form
	\begin{align}	     
	H_{\mathrm{TB}} &= \sum_{i}^N \sum_{\alpha,\beta}  ( E_{i,\alpha} \delta_{\alpha\beta} + \Delta_{i,\alpha\beta}) \, c^{\dagger}_{i\alpha} c_{i\beta}  + \sum_{i}^N \sum_{j \neq i}^N \sum_{\alpha,\beta} t_{i\alpha,j\beta} \, c^{\dagger}_{i\alpha} c_{j\beta}  
	\end{align}
	where $N$ is the number of atoms, $E_{i,\alpha}$ represents on-site energy, $c^{\dagger}_{i\alpha}$ ($c_{i\alpha}$) is the creation (anihilation) operator of the atomic orbital  $\alpha$ on the node $i$. The indices $\alpha$ carry also information about spin (which doubles the number of orbitals). Finally, $\Delta_{i,\alpha\beta}$ accounts for the spin-orbit interaction. 
	In the basis of 
	$
	\ket{\kk,R_n\alpha} =   e^{i \kk \bm{R}_n} \ket{R_n\alpha}
	$
	the Hamiltonian matrix elements can be written
	\begin{align}
	\label{eq:HTB}	     
	H_{n\alpha,m\beta}(\kk) \equiv& \mel{\kk, R_n \alpha}{H_{\mathrm{TB}}}{\kk, R_m\beta} \nonumber \\ =&\ ( E_{n,\alpha} \delta_{\alpha\beta} + \Delta_{n,\alpha\beta}) \, \delta_{nm}  + \ t_{n\alpha,m\beta} \,  e^{i \kk (\bm{R}_m - \bm{R}_n)}\, (1-\delta_{nm}).
	\end{align}	
	
	We perform the calculations using the sp$^3$d$^5$s$^*$ TB Hamiltonian in the nearest neighbors approach, and the $t_{i\alpha,j\beta}$ parameters are expressed in terms of the direction cosines\cite{Slater1954}. The spin-orbit coupling is accounted for taking the elements $\Delta_{n,\alpha\beta}$ between the $p$-shell orbitals\cite{Chadi1977}. We take the set of material parameters from\cite{Jancu1998}.	
	
	To obtain $g(\kk)$ for the Zeeman doublet in the conduction band one need to perform a similar procedure to the case of the $\kp$ model.
	The momentum $P_{i,\alpha\beta}(\kk)$ and the position $R_{i,\alpha\beta}(\kk)$ matrix elements are calculated following Ref.~\cite{LewYanVoon1993}.
	In the first step, this involves
	\begin{align*}
	 &\mel{\kk,R_n\alpha}{\widetilde{P}_{i}}{\kk,R_m\beta} \approx \frac{m_0}{\hbar} \pdv{H_{n\alpha,m\beta}(\kk)}{k_i} = \frac{i m_0}{\hbar} \ (\bm{R}_m - \bm{R}_n)_i \ t_{n\alpha,m\beta} \,  e^{i \kk (\bm{R}_m - \bm{R}_n)},
	\end{align*}	
	then $P_{i,\alpha\beta}(\kk)$ are found using the coefficients $c^{(\alpha)}_m(\kk)$ resulting from the diagonalization of the Hamiltonian. Finally, the $R_{i,\alpha\beta}(\kk)$ are calculated from Eq.~\ref{eq:pos}.

	\subsection{Results}	

	\begin{figure}[tb]
		\begin{center}
			\includegraphics[width=90mm]{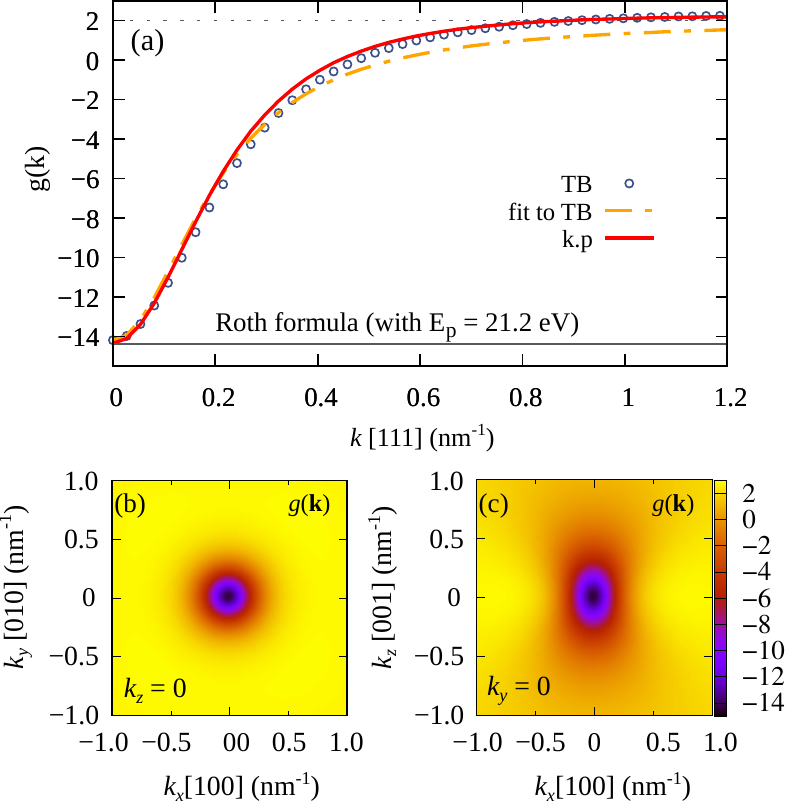}
		\end{center}
		\caption{\label{fig:bulk}\textcolor{gray}({Color online) Electron (conduction band) g-factor in a bulk InAs as a function of $\kk$ along the $[111]$-direction (a). The bottom panels present $g(\kk)$ obtained from $\kp$ model, in cross-section on  the (b) $xy$-, and (c) $xz$ planes. } }
	\end{figure}
		
	We calculated $\kk$-dependent g-factor for the Zeeman doublet in conduction band for the magnetic-field oriented along the $[001]$ direction. 	To avoid spurious solutions in further $\kp$ calculations for nanostructures~\cite{Yong-Xian2010,birner11}, we use a slightly reduced value of $E_{\mathrm{p}} = 21.2$~eV (which is close to $E_{\mathrm{p}} = 21.5$~eV recommended in Ref.~\cite{vurgaftman01}).
	As shown in Fig.~\ref{fig:bulk}(a), the results obtained from the 8-band $\kp$ and the sp$^3$d$^5$s$^*$ TB model are in a good agreement. 	
	At the $\Gamma$ point in the Brillouin zone the g-factor is about $-14.4$ (the $\kp$ model) or $-14.2$ (the TB model). Such a strongly negative value is caused by the spin-orbit interaction, and can be predicted from the Roth-Lax-Zwerdling formula~\cite{Roth1959}. 
	Roth formula however does not contain any $\kk$-dependence, and is generally a poor predictor of g-factors in nanostructures.~\cite{pryor06}

	
	For $\bm{\kk} \neq 0$, the value of g-factor increases which is related to vanishing of the spin-orbit contributions. At high $\kk$, the g-factor tends to its value for free electron $g_0 \approx 2$. However, we note that for larger $\kk$ values $g(\kk)$  can (marginally) exceed $2$ (which is visible in Fig.~\ref{fig:bulk}(a) for $k > 0.9$~nm$^{-1}$). This most likely is a result of several approximations used throughout the calculation, 
	including usage of finite basis~\cite{Boykin2001,pryor06}.
	To further investigate $g(\kk)$  Figs.~\ref{fig:bulk}(b,c) present the cross-sections for the $g(k_x,k_y,0)$ and $g(k_x,0,k_z)$ planes. While, in the first case $g$-factor shows high symmetry, the $xz$ plane exhibit strong anisotropy. This is in agreement to the fact that g-factor is more sensitive to a change of the wave-function in plane perpendicular to the magnetic field than in the field direction~\cite{pryor06,andlauerPhD}. 
	
	Finally, we find that $g(\kk)$ can be quite well approximated the $g(\kk)$ by the following mesoscopic formula
	\begin{equation}
	\label{eq:geff}
	g_{\mathrm{eff}}(\kk) = 2 - \frac{\alpha_0}{1 + \beta_{xy}^2(k^2_x + k^2_y) + \beta_z^2 k^2_z },
	\end{equation}
	where the parameters $\alpha_0 = 16.2$, $\beta_{xy}^2=30$~nm$^2$, $\beta_{z}^2=11$~nm$^2$ were fitted to the TB results, and
	in particular $\alpha_0$ is chosen such that in a limit of 
	$k \to 0$, $g_{\mathrm{eff}} \to 14.2$.
	As shown in Fig.~\ref{fig:bulk}(a), this formula gives a good agreement to the exact results, especially for smaller $k$ values.
	$\beta_{xy}=5.48$~nm and $\beta_{z}=3.32$~nm have a unit of length, in an very loose analogy to the of concept magnetic length~\cite{} for Landau levels (equal to 25.6~nm at 1T). Morever, $\beta_{xy}\neq\beta_{z}$ indicating lack of equivalence of $x$,$y$ and $z$ as the field is applied and oriented along $z$ ([001]) direction.
	
	\section{Nanostructures}
	In the first part of this Section we describe the implementations of magnetic field to nanostructure modeling within the $\kp$ and the TB approaches. Then, we compare the numerical results obtained from both methods for an electron confined in a three dimensional InAs box (cube), as a function of box size (edge length) varying from a single lattice constant, i.e. 0.6 nm to over 120~nm and over 60 million atoms involved in the computations. We show, that these results can be qualitatively well reproduced using an effective model based on the bulk g-factor $g(\kk)$ with virtually no computational cost. We also discuss the effect of symmetry breaking due to underlying crystal lattice, which (in presence of the SO coupling) allows on the mixing of spin configurations in the Zeeman doublet.
	
	\subsection{Eight band k.p model}
	In a standard way, the magnetic field enters the $\kp$ Hamiltonian by the substitution $\kk \rightarrow \kk + (e/\hbar) \bm{A}$. Since, a straightforward implementation leads to gauge dependent results~\cite{Governale1998},  the gauge-invariant scheme was developed~\cite{andlauer08,eissfeller11}. Furthermore, the $\kp$ Hamiltonian is supplemented by the magnetic terms
	\begin{equation*}
	H^{\mathrm{(mag.)}} = \left [ \frac{1}{2}  \mB g_0 \mathcal{S}_z B_z + \mB \hat{\mathcal{L}}_z B_z + \cp \right ] + \widetilde{H}^{\mathrm{(r)}},
	\end{equation*}
	where\cite{Winkler2003}
	\begin{align}
	\label{eq:kp_sz}
	\mathcal{S}_z =&  (\sigma_z)_{\mathrm{6c6c}} + \frac{2}{3}  (J_z)_{\mathrm{8v8v}} - \frac{1}{3}  (\sigma_z)_{\mathrm{7v7v}} - 2 (T_z)_{\mathrm{7v8v}}  - 2 (T^\dagger_z)_{\mathrm{8v7v}} 
	\end{align}
	is related to spin. The band angular momentum (which was neglected in the $\kp$ calculations in the previous Section) is represented by
	\begin{align}
	\label{eq:kp_lz}
	\mathcal{L}_z =&  \frac{2}{3}  (J_z)_{\mathrm{8v8v}} + \frac{2}{3}  (\sigma_z)_{\mathrm{7v7v}} + (T_z)_{\mathrm{7v8v}}  + (T^\dagger_z)_{\mathrm{8v7v}}. 
	\end{align}	
	Finally, the $ \widetilde{H}^{\mathrm{(r)}}$  describes contributions from the remote bands, which are not explicitly included in a given $\kp$ model~\cite{eissfeller12}. As the calculated electron band-edge g-factor ($g = -14.4$) is already close to the experimental value ($g = -14.7$~\cite{Konopka1967}), we neglect these remote band contributions.

	\subsection{Tight binding model}	
	The magnetic field is implemented to the TB Hamiltonian by the standard Peierls substitution~\cite{Graf1995,Boykin2001} 
	\begin{align*}
		t_{i\alpha,j\beta} \rightarrow t_{i\alpha,j\beta} \, e^{i \theta_{ij}},
	\end{align*}	
	with a phase given by
	\begin{equation*}
	\theta_{ij} = \frac{e}{\hbar} \int_{\bm{R}_i}^{\bm{R}_j} \bm{A}(\rr) d\bm{l} \approx \frac{e}{\hbar}  \frac{\bm{A}(\bm{R_i}) + \bm{A}(\bm{R_j})}{2} \cdot (\bm{R_j} - \bm{R_i}).
	\end{equation*}
	Assuming constant magnetic field and a symmetric gauge, one can obtain\cite{Vogl2002}
	\begin{equation*}
	\theta_{ij} = \frac{e}{2 \hbar} \bm{B} \cdot (\bm{R}_i \cross \bm{R}_j).
	\end{equation*}
	The contribution from the spin is accounted for via the on-site terms~\cite{Graf1995}
	\begin{align}
	\label{eq:tb_sz}
	H^{\mathrm{(spin)}}_{n\alpha,m\beta}(\kk)  &=  \frac{1}{2} \mB g_0 \left [ (\sigma_{z})_{\alpha\beta} B_z + \cp \right ] \delta_{nm},
	\end{align}
	where $(\sigma_{i})_{\alpha\beta}$ are matrix elements of the Pauli matrices in the basis of atomic orbitals (note that indices $\alpha$ contains also information about the spin). 
	This can be viewed as a tight-binding analogue of Eq.~\ref{eq:kp_sz}.
	
	For nanostructures, in order to avoid any spurious states in the energy range of interest, the dangling bonds of surface atoms are passivated according to well established approach of Ref.~\cite{Lee2004}.
	We found that tight-binding results somewhat depend on the choice of the dangling bond shift, however the trends obtained with different shifts are very similar, in agreement with conclusions of Ref.~\cite{Tadjine2017}.
	Strain effects and piezoelectricty are not present in a system due to use of a single chemical compound.
	For sake of comparison with the $\kp$ and the effective model, any surface reconstruction or presence of image charges are also neglected.

	\begin{figure}[tb]
		\begin{center}
			\includegraphics[width=85mm]{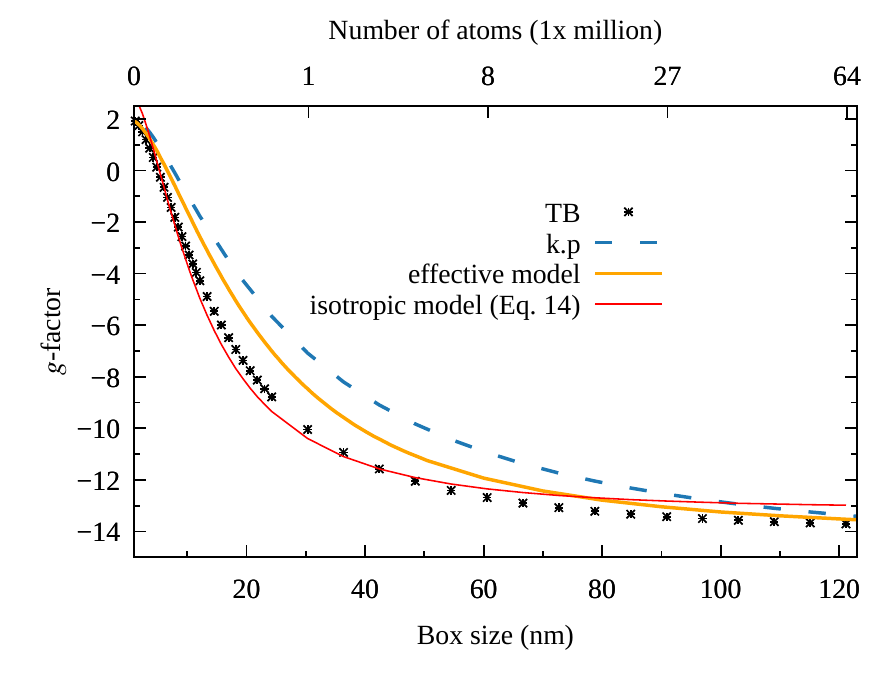}
		\end{center}
		\caption{\label{fig:box_Td}\textcolor{gray}({Color online) The electron (conduction band) g-factor in the same-anion-terminated InAs cubic box of $T_{\mathrm{d}}$ symmetry as a function of box size, and calculated using three different approaches. Corresponding number of atoms used in atomistic tight-binding calculations is shown in the upper axis. See text for details.} }
	\end{figure}
	\subsection{Results: electron in a box}		

    In the following, we calculate the electron g-factor for nano-size InAs cubic box, and by varying box size simultaneously in all three dimensions. We start with a box size of one lattice constant edge length, corresponding to very high quantum confinement, and we progress through intermediate dimensions to very large InAs boxes, with dimensions extending well over 100 nm aiming for near bulk-like properties.
    The largest of considered box (200$\times$200$\times$200 in lattice constant units) has edge length equal to 120 nm, corresponding to $64.2$ million atoms in the calculation. In this case, the ground electron state energy is 0.422~eV, thus only 3 meV larger than the bulk limit of 0.418 eV.
    We note as well, in largest considered cases, the tight-binding calculation was performed on parallel computer cluster with 192 computational cores, and MPI-parallelized Lanczos solver, taking approximately 12 hours to find several lowest electronic states, and
    the time of the computation scaling proportionally with number of atoms for smaller systems~\cite{Rozanski2016}.

    The numerical results are presented in Fig.~\ref{fig:box_Td}, where 
    the values of the g-factor are taken from the energy differences between the two lowest states $g = (E_2 - E_1)/(\mB B) $, calculated at $B = 1$~T. The sign is determined from the spin orientations of states, which are represented by the averages $\expval{S_z}$. 
    
    The results obtained from the TB and $\kp$ models provide qualitatively similar $g$-factor dependence.
    In both cases, for a very strong confinement the value of the $g$-factor is close to the free electron case ($g_0$). 
    Moreover, in both cases the increasing size brings $g$-factor to the bulk value in a similar fashion. 
    Yet, there is a notable difference between both approaches, with $\kp$ systematically reporting smaller $g$-factor magnitudes. However, for a fairness of comparison, quantitative differences of that kind are expected, since we compare 20-band atomistic model (including $d$-orbitals), and 8-band method based on continuous media approximation. Both method report somewhat different bulk band structures, utilize different of treatment of the cube surface, boundary conditions etc.  
    
    We also performed the simulations in a simple effective model which relies on the $g_\mathrm{eff}(\kk)$ for bulk InAs (see Eq.~\ref{eq:geff}) and on the Fourier transform of the wave-function. For sake of simplicity, and computational efficiency, we assume that the electron ground state is build from the bulk conduction band and having the envelope of the three-dimensional infinite well (with $L \times L \times L$ size)
    \begin{equation}
    \label{eq:simpleenv}
    \Psi(\rr) = \sqrt{\frac{8}{L^3}} \sin(\frac{x\pi}{L}) \sin(\frac{y\pi}{L}) \sin(\frac{z\pi}{L}).
    \end{equation}
    Utilizing the separation of variables, the envelope representation in the $\kk$ can be expressed by
    \begin{equation*}
    \Phi(\kk) = \phi_x(k_x) \phi_y(k_y) \phi_z(k_z),
    \end{equation*}
    where~\cite{Majernik1997}
    \begin{equation*}
    \phi_n(k_i) = \sqrt{\frac{L}{\pi}} \frac{2 \pi}{\pi+k_i L} \frac{\sin\left (  \frac{\pi-k_i L }{2}  \right )}{{\pi-k_i L }}.
    \end{equation*}
    Then, the effective g-factor can be calculated from
    \begin{equation*}
    g_{\mathrm{box}}(L) = \int_{-\infty}^{\infty} \abs{\Phi(\kk)}^2 g_\mathrm{eff}(\kk)  \dd \kk.
    \end{equation*}
    As shown in Fig.~\ref{fig:box_Td}, this surprisingly simple, effective model provided results which are in-between $\kp$ approach, and the multi-million atom simulations obtained using the TB model.
    By performing several numerical tests, we found that the
    $g_{\mathrm{box}}(L)$ differs tight-binding prediction 
    mostly due to oversimplified Eq.~\ref{eq:simpleenv} assuming hard-wall boundary conditions, 
    whereas approximation of $g_\mathrm{eff}$ plays a lesser role.
    
    It is known that electron g-factor for any nanostructure can be decomposed into an isotropic part (depending on the effective band gap) and the surface part which depends on the shape of the structure~\cite{Tadjine2017}. The first part can be expressed by~\cite{Tadjine2017}
    \begin{equation}
    g(E) = \tilde{g}_0 - \frac{E^2_0}{E^2},
    \end{equation}
    where $E$ is the effective band gap, and $\tilde{g}_0$, $E_0$ are parameters which can be fitted for a given material. We used this formula for calculations in the box (the red line in Fig.~\ref{fig:box_Td}). Although for the  parameters $\tilde{g}_0 = 3.09$ and $E_0 = 1.69$~eV (which we fitted to the TB results) we obtain a very good agreement in middle part of the plot, we cannot simultaneously reproduce the values at the boundaries (in the limit of a very small/large box). This suggest, that in the considered system the part of the $g$-factor related to the surface plays important role.

	\begin{figure}[tb]
		\begin{center}
			\includegraphics[width=85mm]{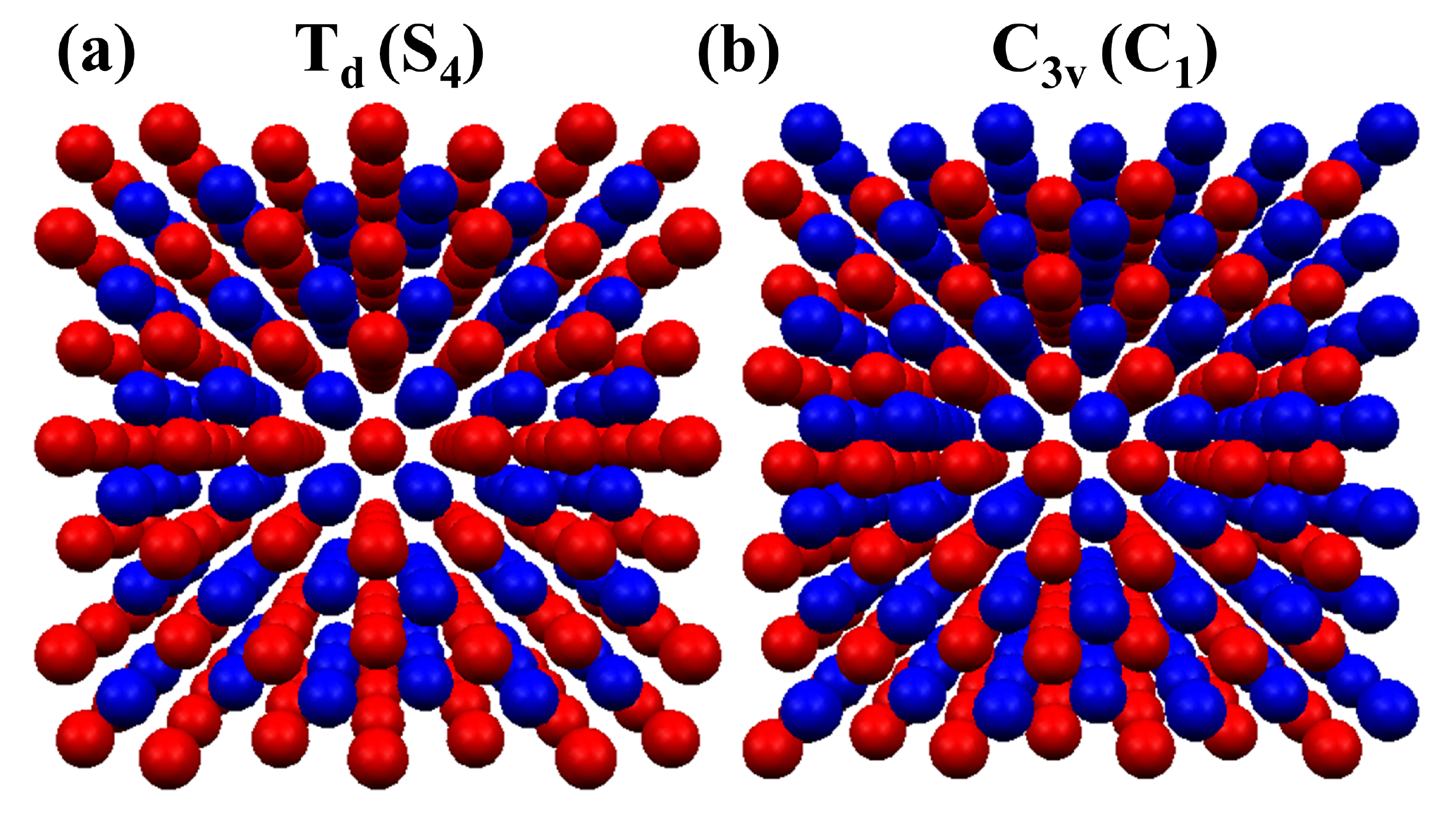}
		\end{center}
		\caption{\label{fig:boxes_schematic}\textcolor{gray}({Color online) The schematic view of two cubic boxes: (a) the same ion-terminated corresponding to $S_4$ symmetry and (b) mixed-anion-cation-terminated leading to a lower $C_1$ symmetry.
		Red color corresponds to anions (arsenic) and blue color marks cation (indium).}}
	\end{figure}		
\subsection{Role of atomistic symmetry}
    So far we have studied nanostructures without analyzing the role of underlying crystal lattice~\cite{Singh2009,Zielinski2013a}.
    We note that nanostructures analyzed in Fig.~\ref{fig:box_Td} were 
    cubic boxes cut from zinc-blende lattice, obtained by terminating box sides with the same ionic species Fig.~\ref{fig:boxes_schematic}(a). Alternatively such geometry can be obtained by choosing the box geometric center to be placed on one of ions. In case of Fig.~\ref{fig:box_Td}(a) this center is placed on anion (marked as red spheres), and results discussed so far where obtained for this particular choice.
    
    Such atomic arrangement leads to an overall $T_{\mathrm{d}}$ (tetrahedral) symmetry of a nanostructure, which may not be apparent from inspection of Fig.~\ref{fig:boxes_schematic}(a), however it was additionally verified with Jmol~\cite{Jmol} and Chemcraft~\cite{Chemcraft} tools allowing for point-group symmetry determination.
	\begin{figure}[tb]
		\begin{center}
			\includegraphics[width=85mm]{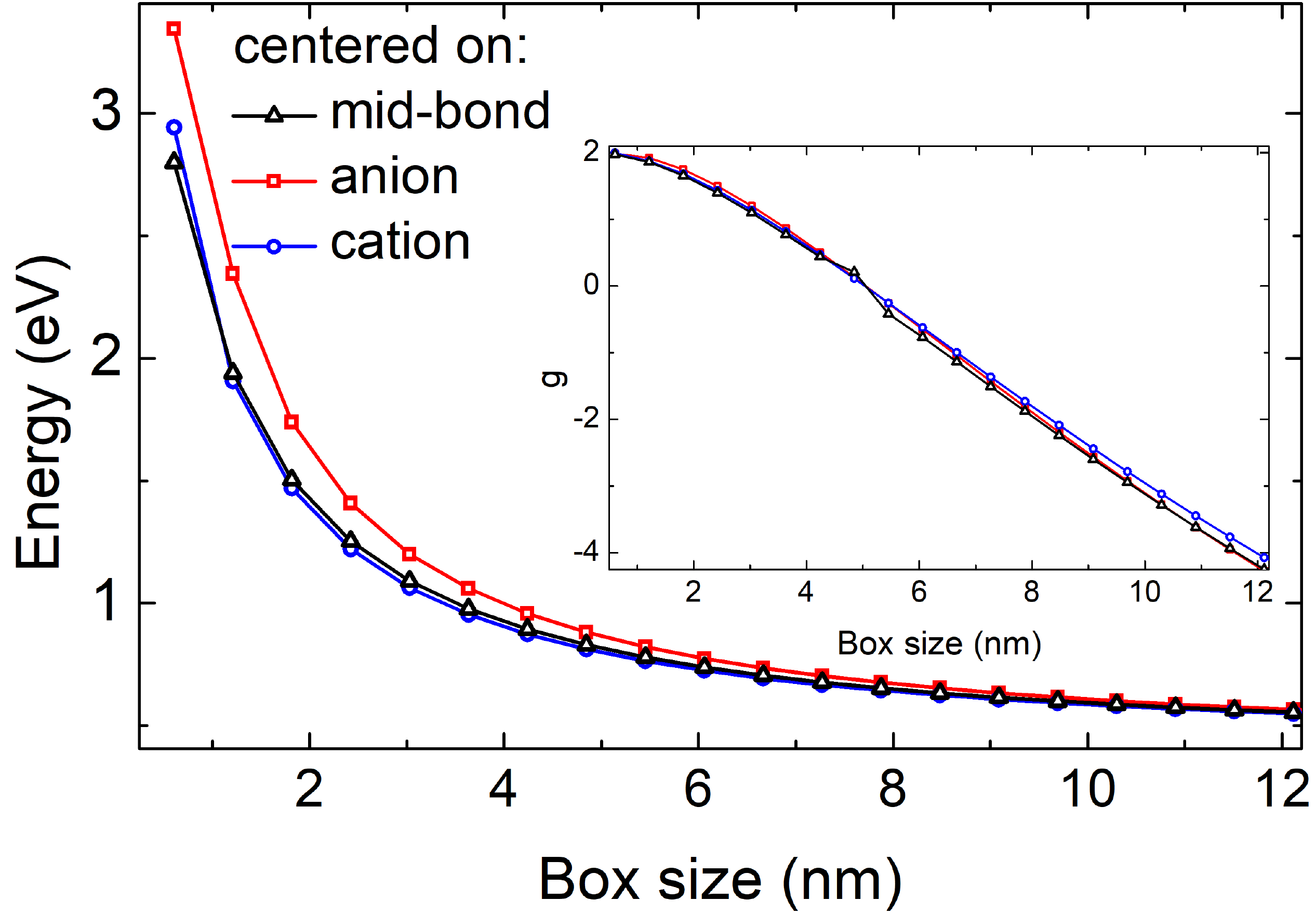}
		\end{center}
		\caption{\label{fig:anion_cation}\textcolor{gray}({Color online) 
		The ground electron state energy as a function of box size for different choice of origin, i.e. centered on anion, cation, and mid-bond respectively. Inset shows corresponding values of electron $g$-factor.
		See text for details.}}
	\end{figure}		    
    Moreover, for the same $T_{\mathrm{d}}$ symmetry, one can choose either cation or anion as an origin.
    Centering the system on the cation effectively corresponds of replacing anions with cations [or red with blue atoms in Fig.~\ref{fig:boxes_schematic}(a)].
    In both cases (whether centered on anion or cation) the overall symmetry is $T_{\mathrm{d}}$ provided that box is terminated consistently with one ionic species only.
    As we consider magnetic field oriented in the [$001$] direction, the field effectively lowers the symmetry of the system from the $T_{\mathrm{d}}$ to the $S_4$ point group, therefore the tight-binding results in Fig.~\ref{fig:box_Td} corresponded to $S_4$ symmetry.
    
    Even though not altering the symmetry, switching indium atomic positions with arsenics, may substantially affect single particles energies, however only for smaller boxes, as shown in Fig.~\ref{fig:anion_cation}. These differences are related to different ionic composition~\cite{Jancu1998} of ground electron states, since nanostructure shown in  Fig.~\ref{fig:boxes_schematic}(a) will have different number of cation and anion depending on the choice of origin. 
    For the anion-centered case (and anions terminating the surface), with overall number of anions is larger than cations. 
    For the cation-centered case the situation is exactly opposite.
    Centering the origin on the anion or cation will thus not alter symmetry, but will change the overall ionic ``stoichiometry''.
    Nonetheless, despite substantial differences in single particle energies for smaller boxes, anion-cation centering issues have a rather small effect on the final $g$-factor dependence (inset on Fig.~\ref{fig:anion_cation}).
    Interestingly, a small difference in $g$-factor values between anion and cation centered cases can be observed even for larger boxes, again due to difference in stoichiometry, however it decays with system size, and for the edge length of 60 nm (100$\times$100$\times$100 box; including 4$\times 10^6$ cations and 4.06$\times 10^6$ anions) the $g$-factor varies by about 1\% with respect to cation-centered case (with 4.06$\times 10^6$ cation and 4$\times 10^6$ anions).
    
    Another possible choice of crystal lattice arrangement with respect to box shape is presented in Fig.~\ref{fig:boxes_schematic}(b), with corresponding tight-binding results shown for comparison in Fig.~\ref{fig:anion_cation}. This particular case is given by a seemingly unimportant shift of the box origin by half of a bond length, i.e. placing it in the exactly between anion and cation.
    Such choice also leads to termination of box sides in a mixed anion-cation fashion as apparent from Fig.~\ref{fig:boxes_schematic}(b).
    Moreover mid-bond case corresponding to exactly equal number of cations and anion, for all considered dimensions. 
    Although there is not strict inversion symmetry in zinc-blende lattice, further replacing anions with cations in Fig.~\ref{fig:boxes_schematic}(b) leads to virtually the same single particle spectra (with $\mu$eV differences), therefore it is not considered here.
    
    However, despite ideal stoichiometry, a closer inspection reveals that anion/cation mixing at the boundaries leads to an overall symmetry reduction to $C_{3v}$, and with further reduction of symmetry to $C_1$, when the magnetic field along z-axis is applied.
    Yet, as seen in Fig.~\ref{fig:anion_cation} there is virtually no difference to previous results, with the exception of near-zero $g$-factor values.
    \begin{figure}[tb]
		\begin{center}
			\includegraphics[width=90mm]{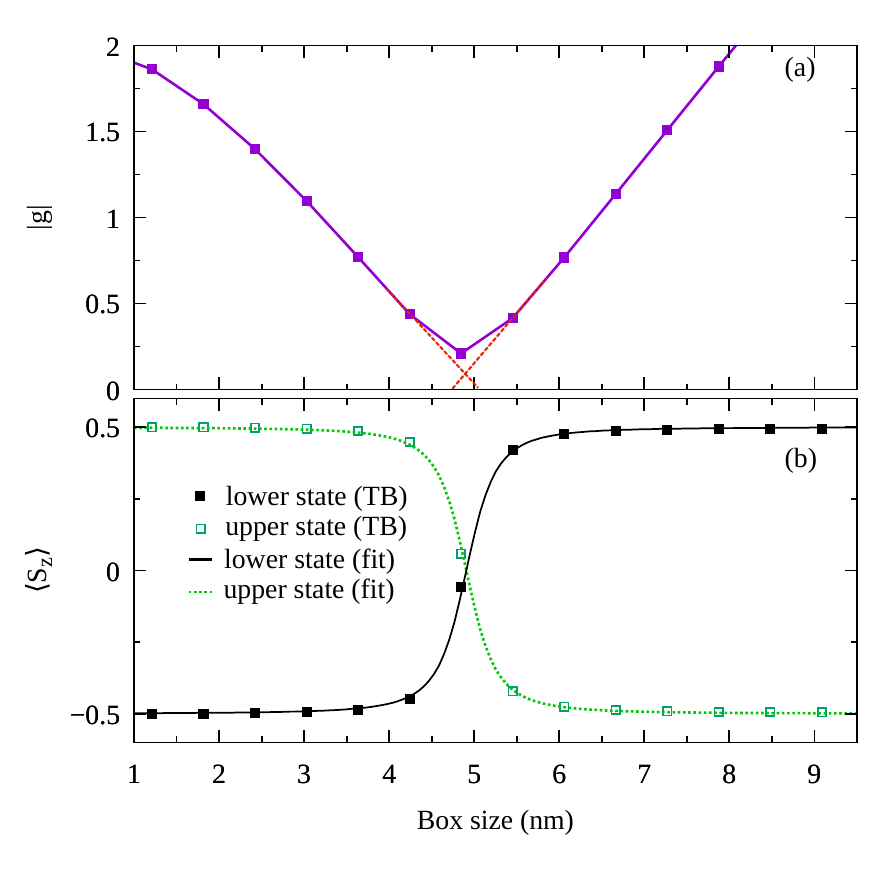}
		\end{center}
		\caption{\label{fig:box_C1}\textcolor{gray}({Color online) (a) The absolute value of the electron (conduction band) g-factor in a mixed-anion-cation-terminated InAs cubic box of $C_{\mathrm{1}}$ symmetry. The red dashed lines are a guide for eye. (b) Average value of the spin $z$-th component for the states in the Zeeman doublet. } }
	\end{figure}
    In the previous case of the high symmetry system (whether anion or cation centered), the $g$-factor value smoothly varied with the box size and crossed $0$ (which simply corresponded to a swap of the states in the Zeeman doublet).  In contrast, for the low symmetry box the $g$-factor changes its sign avoiding 0 (while the absolute value remains continuous). This is better illustrated on  Fig.~\ref{fig:box_C1}(a) where now absolute value of $g$-factor is now presented for mid-bond (low symmetry) case only, and with a magnification of close to g-zero regions. This peculiar feature in $g$-factor dependence corresponds to an anti-crossing between the states in the Zeeman doublet, which is reflected in the values of $\expval{S_z}$ [Fig.~\ref{fig:box_C1}(b)]. This behavior is related to the symmetry selection rules. According to the group theory, two states can be mixed (which manifests in an anti-crossing of their energy levels) if they belong to the same irreducible representation~\cite{Bir1974,Dresselhaus2010}. In the case of the same-ion-terminated box, the symmetry of the system is described by the $S_4$ point group (which in presence of the spin-orbit coupling needs to be a double group). Then, the two lowest electron states belong to different irreducible representations and coupling between them is prohibited. In contrast, for the mixed-anion-cation-terminated box, the symmetry is reduced to the $C_1$ (which including spin is also a double group). In this case, the states in the Zeeman doublet belong to the same representation. Hence, their energies exhibit anti-crossing and spin configurations mixes (which is in fact caused by the spin-orbit coupling).
    This effect will thus happen for any low atomic arrangement, and is not limited to a particular choice presented in Fig.~\ref{fig:boxes_schematic}(b), although application of magnetic field exactly along high symmetry crystal axes, such as [111] (diagonal axis on Fig.~\ref{fig:boxes_schematic}(b)) would restore high-symmetry, and allow for electron level crossing rather than anti-crossing.
    However, no nanostructure can ever be grown to have an ideal atomistic symmetry as systems presented in Fig.~\ref{fig:boxes_schematic}. In fact, removal of just a single atom can break the overall symmetry and thus will lead to level anti-crossing. The effect of anti-crossing could be also viewed as a presence of off-diagonal terms in the g-tensor. It is known that g-tensors in low symmetry systems (the $C_1$ group without magnetic field) have 9 independent components~\cite{Ivchenko1997}, which cannot be reduced to the diagonal form for any magnetic field direction. 
    
    \section{Conclusions}
    \label{sec:concl}
    In summary, with atomistic tight-binding method, and continuous media approximation $\kp$ approach we have studied Land\'e $g$-factor $\kk$-dependence for InAs. Based on that, we proposed a mesoscopic model with three effective parameters only, and further we compared this model with multi-million atom tight-binding calculations, for a cubic nanostructure with dimensions varying from single to 120 nanometers and number of atoms reaching 64 million. 
    Despite its simplicity the mesoscopic model shows a good qualitative dependence with actual results ranging between that of the tight-binding and the $\kp$ at virtually no computation cost.
    
    Further, we have inspected nanostructure dimensions corresponding to near-zero $g$-factor, and found that depending on a detail of atomic arrangement electron spin states can undergo an anti-crossing as a function of system size. 
    The effect occurs despite high-shape symmetry, and is not related to cation-anion stoichmetric imbalances, but due to low overall symmetry being a result of cubic shape imposed on underlying zinc-blende lattice. 
    Our results therefore emphasize the key role of symmetry in nanostructures, and show inherent limits to $g$-factor tuning, especially important for applications involving near-zero  $g$-factor values.

	\section*{Acknowledgments}
	The authors acknowledge funding from the Polish National Science Centre (NCN) under Grant
	No. 2016/23/G/ST3/04324 (K.G.) and the support from the Polish National Science Centre based on Decision No. 2018/31/B/ST3/01415 (M.Z.).
	Calculations have been carried out using
	resources provided by Wroclaw Centre
	for Networking and Supercomputing (\url
	{http://wcss.pl}), Grant No.~203. We are also grateful to Micha{\l} Gawe{\l}czyk for valuable discussions.

	\section*{Author Contributions}
	K.G. performed analytical calculations and numerical simulations for the $\kp$ model and for the tight-binding (the bulk case only). M.Z. performed the TB numerical simulations for the box.  
	All authors contributed equally to analyzing and interpreting the results, as well as to writing and reviewing the manuscript.

	\bibliography{abbr,../library.bib}
	
\end{document}